\documentclass[12pt]{article}
\usepackage[utf8]{inputenc}
\usepackage[english]{babel}
\usepackage{amssymb}
\usepackage{amsmath}
\usepackage{graphicx}
\usepackage[small,sc,center]{titlesec}
\usepackage{indentfirst}
\usepackage{siunitx}

\newcommand{\nc}{\newcommand}
\nc{\itp}{i_\mathrm{TP}}
\nc{\lhe}{L_\mathrm{He}}
\nc{\mzams}{M_\mathrm{ZAMS}}
\nc{\rph}{R_\mathrm{ph}}
\nc{\teff}{T_\mathrm{eff}}
\nc{\tev}{t_\mathrm{ev}}

\begin{document}

\begin{center}
\textbf{Pulsations of intermediate--mass stars on the asymptotic giant branch}

\vskip 3mm
\textbf{Yu. A. Fadeyev\footnote{E--mail: fadeyev@inasan.ru}}

\textit{Institute of Astronomy, Russian Academy of Sciences, Pyatnitskaya ul. 48, Moscow, 119017 Russia} \\

Received February 20, 2017
\end{center}

\textbf{Abstract} ---
Evolutionary tracks from the zero age main sequence to the asymptotic giant branch
were computed for stars with initial masses $2M_\odot \le \mzams \le 5M_\odot$ and
metallicity $Z=0.02$.
Some models of evolutionary sequences were used as initial conditions for equations
of radiation hydrodynamics and turbulent convection describing radial stellar
pulsations.
The early asymptotic giant branch stars are shown to pulsate in the fundamental mode
with periods $30~\mathrm{day}\lesssim\Pi\lesssim 400~\mathrm{day}$.
The rate of period change gradually increases as the star evolves but is too small
to be detected ($\dot\Pi/\Pi < 10^{-5}~\text{yr}^{-1}$).
Pulsation properties of thermally pulsing AGB stars are investigated on time intervals
comprising 17 thermal pulses for evolutionary sequences with initial masses
$\mzams=2M_\odot$ and $3M_\odot$ and 6 thermal pulses for $\mzams=4M_\odot$ and $5M_\odot$.
Stars with initial masses $\mzams\le 3M_\odot$ pulsate either in the fundamental mode or
in the first overtone, whereas more massive red giants ($\mzams\ge 4M_\odot$) pulsate in
the fundamental mode with periods $\Pi\lesssim 10^3$ day.
Most rapid pulsation period change with rate
$-0.02\lesssim\dot\Pi/\Pi\lesssim -0.01~\text{yr}^{-1}$
occurs during decrease of the surface luminosity after the maximum of
the luminosity in the helium shell source.
The rate of subsequent increase of the period is
$\dot\Pi/\Pi\lesssim 5\times 10^{-3}~\text{yr}^{-1}$.


\newpage
\section{introduction}

Miras belong to long--period late--type variable stars with visual light amplitude
greater than 2 magnitudes and pulsation period as long as $10^3$ day.
Light variations of Mira--type variables are not strictly periodic and at least three
kinds of period instability are distinguished.
First, random cycle--to--cycle fluctuations of the pulsation period
(Eddington, Plakidis, 1929) seem to be a common property of Miras,
the nature of fluctuations remaining unknown (Percy, Colivas, 1999).
Second, some Miras show secular change (either increase or decrease) of the period.
This property of Miras has been firstly reported by Eddington and Plakidis (1929)
who found the secular period change in $o$~Cet and $\chi$~Cyg.
The fraction of Miras with detected long-term period change is currently estimated
in the range from one to a few per cent (Templeton et al., 2005).
The secular period change of Miras is due to variations of the stellar radius and
surface luminosity accompanying the thermal pulse in the helium shell source
(Wood, Zarro, 1981).
Third, as many as a quarter of Miras and semiregular (SR) variables exhibit the
long secondary periods (LSP) by an order of magnitude longer than the
fundamental mode period (Wood et al., 2004).
The physical origin of the secondary periodicity is unclear.

Discovery of absorption lines of the radioactive element technetium with
half--life time $\lesssim 4.2\times 10^6$ yr in atmospheres of Miras
of spectral types S and C (Merrill, 1952; Busso et al., 1999) indicates that 
these variables are asymptotic giant branch (AGB) stars.
Secular changes of the pulsation period provide further evidence for Miras
as AGB stars.
In particular, Wood and Zarro (1981) have shown that the rates of period change
in Miras R Aql, W Dra and R Hya are consistent with results of evolutionary
computations done for thermally pulsing AGB stars.
However, one should be noted that the authors did not investigate pulsational
instability of red giants and the theoretical estimates of the period change rate
they obtained in assumption that Miras are the first overtone pulsators.

The pulsation mode of galactic Miras is still disputable.
Direct radius measurements of brightest Miras by Weiner (2004) and Perrin et al. (2004)
show that observed stars pulsate in the fundamental mode.
The same conclusion was made by Ya'ari and Tuchman (1999) who computed
hydrodynamic models of red giants with masses $1M_\odot\le M\le 1.3M_\odot$
and luminosities $2400L_\odot\le L\le 4000L_\odot$.
At the same time multi--wavelength angular diameter measurements of nearby Miras
done by Tuthill et al. (1994) and Haniff et al. (1995) show that Miras are
the first--overtone pulsators.
Observations of Miras in the Large Magellanic Cloud (LMC) also are contradictory.
Wood and Sebo (1996) and Wood et al. (1999) have shown that Miras in LMC are
the fundamental mode pulsators, whereas Ita et al. (2004) provided evidences
favoring both fundamental mode and first overtone pulsations.
The same conclusion was drawn from photometric monitoring of long period variable
stars in dwarf galaxies NGC~147 and NGC~185 (Lorenz, 2011).

One of the causes of the above contradiction is that there are still significant
uncertainties in pulsation properties of late--type giant stars.
Attempts to study nonlinear pulsation models of red giants were undertaken
by a number of authors but effects of turbulent pressure and turbulent
viscosity were ignored or were not considered in detail
(Keeley, 1970; Smith, Rose, 1972; Wood, 1974; Tuchman et al., 1978; 1979;
Perl, Tuchman, 1990; Cox, Ostlie, 1993; Hofmann et al., 1998;
Ya'ari, Tuchman, 1999).
First hydrodynamic computations for red giant models based on sufficiently
realistic treatment of turbulent convection were done by Olivier and Wood (2005).
They employed the convection model of Kuhfu\ss (1986) and showed that
hydrodynamic models taking into account interaction between gas
flow and turbulence via turbulent viscosity are in a better agreement with
observations than those of previous works.
Unfortunately, this study was restricted by exploratory computations of five
models with masses $1M_\odot$ and $5M_\odot$.
Later this approach was used by Kamath et al. (2010) for computation of
several low--mass AGB star models with composition of the Magellanic Cloud
clusters NGC~1978 and NGC~419.

The present work is aimed at determining the pulsation properties
(i.e. the pulsation mode and the pulsation period) as a function of evolutionary
time $\tev$ for AGB stars with initial masses $2M_\odot\le\mzams\le 5M_\odot$
and initial relative mass fractions of helium and heavier elements
$Y=0.28$ and $Z=0.02$, respectively.
Results of computations are compared with observations of Mira--type variables
having most reliable estimates of the period change rate.
Below we extend our earlier results obtained for the first ten thermal pulses
of the population~I AGB star with initial mass $\mzams=3M_\odot$ (Fadeyev, 2016).

\section{methods of computation}

In this study we solved the Cauchy problem for equations of one--dimensional radiation
hydrodynamics and turbulent convection with initial conditions corresponding
to hydrostatic equilibrium.
Initial conditions were determined from stellar evolution computations.
Below we briefly describe the methods employed to calculate the stellar evolutionary
models and nonlinear stellar pulsations.

\subsection{stellar evolution}

Evolutionary sequences were computed with the MESA code version 8845
(Paxton et al., 2011; 2013; 2015).
Evolution of isotopic abundances and the rate of energy generation were
computed for the included reaction network consisting of 24 isotopes
from ${}^{1}\mathrm{H}$ to ${}^{24}\mathrm{Mg}$ coupled by 42 reactions.
Convection was treated through the mixing length theory by B\"ohm--Vitense (1958)
with a mixing length to pressure scale height ratio
$\alpha_\mathrm{MLT} = \ell/H_\mathrm{P} = 1.8$.
Mixing beyond convective boundaries was treated with an exponentially decaying
diffusion coefficient (Herwig, 2000)
\begin{equation}
 D_\mathrm{ov}(z) = D_0 \exp\left(-\frac{2z}{f H_\mathrm{P}}\right) ,
\end{equation}
where
$H_\mathrm{P}$ is the local pressure scale height,
$D_0$ is the convective diffusion coefficient (Langer, 1985) at
the layer inside the convection zone and distant from the Schwarzschild
boundary by $0.04H_\mathrm{P}$,
$z$ is the radial distance from this layer.
For all convection zones an adjustable parameter is $f=0.016$.

The mass loss rate $\dot M$
during the red giant evolutionary stage preceding core helium ignition
was evaluated in units of $M_\odot\:\mathrm{yr}^{-1}$
following the prescription of Reimers (1975)
\begin{equation}
\dot M_\mathrm{R} = 4\times 10^{-13} \eta_\mathrm{R} (L/L_\odot)
                    (R/R_\odot) (M/M_\odot)^{-1} .
\end{equation}
For the AGB stage  we followed the prescription of Bloecker (1995)
\begin{equation}
\dot M_\mathrm{B} = 4.83\times 10^{-9} \eta_\mathrm{B} (M/M_\odot)^{-2.1}
                    (L/L_\odot)^{2.7} \dot M_\mathrm{R} .
\end{equation}
All the evolutionary sequences were computed with mass loss parameters
$\eta_\mathrm{R}=0.5$ and $\eta_\mathrm{B}=0.05$.

\subsection{nonlinear stellar pulsations}

Equations of one--dimensional radiation hydrodynamics were solved
together with transport equations for time--dependent convection in
spherically--symmetric geometry (Kuhfu\ss, 1986).
The system of equations and the adopted parameters of the convection
model are described in our previous paper (Fadeyev, 2015).

Hydrodynamic computations were carried out on the Lagrangian grid
consisting of $500\le N\le 2000$ mass zones.
The larger number of mass zones was found to provide a better convergence
of the Newton--Raphson iteration procedure in calculation of more luminous
hydrodynamic models.
The mass intervals of the Lagrangian grid increase geometrically inwards
but 100 to 400 innermost intervals geometrically decrease.
Such an approach allowed us to significantly reduce approximation errors of
difference equations near the bottom of the convection zone where the pressure
and the temperature abruptly increase inwards.

The inner boundary of the hydrodynamic model is set at the layer with radius
$r_0\lesssim 0.05R$, where $R$ is the outer radius of the hydrostatically
quilibrium evolutionary model.
Both the radius and the luminosity at the inner boundary of the
hydrodynamic model are assumed to be time--independent.
Initial values of the variables defined at the zone centers and the zone
interfaces of the Lagrangian grid were obtained by means of interpolation
of the evolutionary model data (the radius $r$, the luminosity $L_r$,
the total pressure $P$ and the temperature $T$ as a function of mass
coordinate $M_r$).
Interpolation errors played the role of initial hydrodynamic perturbations.

Solution of the equations of hydrodynamics with respect to time $t$
was accompanied by evaluation of the kinetic energy of pulsation motions
\begin{equation}
E_\mathrm{K}(t) = \frac{1}{2}\sum\limits_{j=1}^N \Delta M_j U(t)_j ^2 ,
\end{equation}
where $U(t)_j$ is the gas flow velocity at the $j$--th zone interface
and $\Delta M_{j-1/2}$ is the mass of the interval between $(j-1)$--th and
$j$--th zone interfaces.
Radial pulsations of red giants are described by standing waves to a good
accuracy, so that the kinetic energy changes during the pulsation cycle by
several orders of magnitude.
Therefore, to determine the pulsational instability we evaluated the
maximal over the pulsation cycle values $E_\mathrm{K,max}$.
In the case of pulsational stability $E_\mathrm{K,max}$ exponentaially
reduces with time $t$.
To be confident that all pulsation modes are stable the computations
were continued until the kinetic energy decreases by several orders
of magnitude in comparison with the energy of initial perturbation.

In the case of pulsational instability the stage of exponential growth of
the kinetic energy is followed by the stage of the limit cycle when
$E_\mathrm{K,max}$ becomes time--independent.
It should be noted that the limit--cycle condition is fulfilled to
a good accuracy only for less luminous red giants.
Increase of luminosity is accompanied by growth of the amplitude of limit cycle
oscillations so that models of more luminous stars show significant
nonlinear effects responsible for amplitude and period irregularities.

After the cease of the growth of $E_\mathrm{K,max}$ the hydrodynamic computations
were carried out for each model on the time interval comprising from $10^2$
to $10^3$ pulsation cycles in order to evaluate the pulsation period $\Pi$
and the mean radius of the photosphere $\langle\rph\rangle$.
To this end we used a discrete Fourier tranform of the temporal
dependences of $E_\mathrm{K}(t)$ and $\rph(t)$.
The pulsation constant $Q$ was evaluated from the period--mean density
relation
\begin{equation}
\Pi = Q \frac{(\langle\rph\rangle/R_\odot)^{3/2}}{(M/M_\odot)^{1/2}} .
\end{equation}

\section{results of computations}

The life--time of the red giant evolving through the early asymptotic giant
branch (eAGB) stage is several times longer than the following stage of the
thermally pulsing helium shell source (TP--AGB).
Moreover, the eAGB star is in hydrostatic and thermal equilibrium whereas
thermal pulses during the TP--AGB stage lead to thermal imbalance
in the stellar envelope.
Therefore below we discuss the pulsational properties of eAGB and TP--AGB stars
separately.

\subsection{eAGB stars}

\begin{figure}
\centerline{\includegraphics[width=16cm]{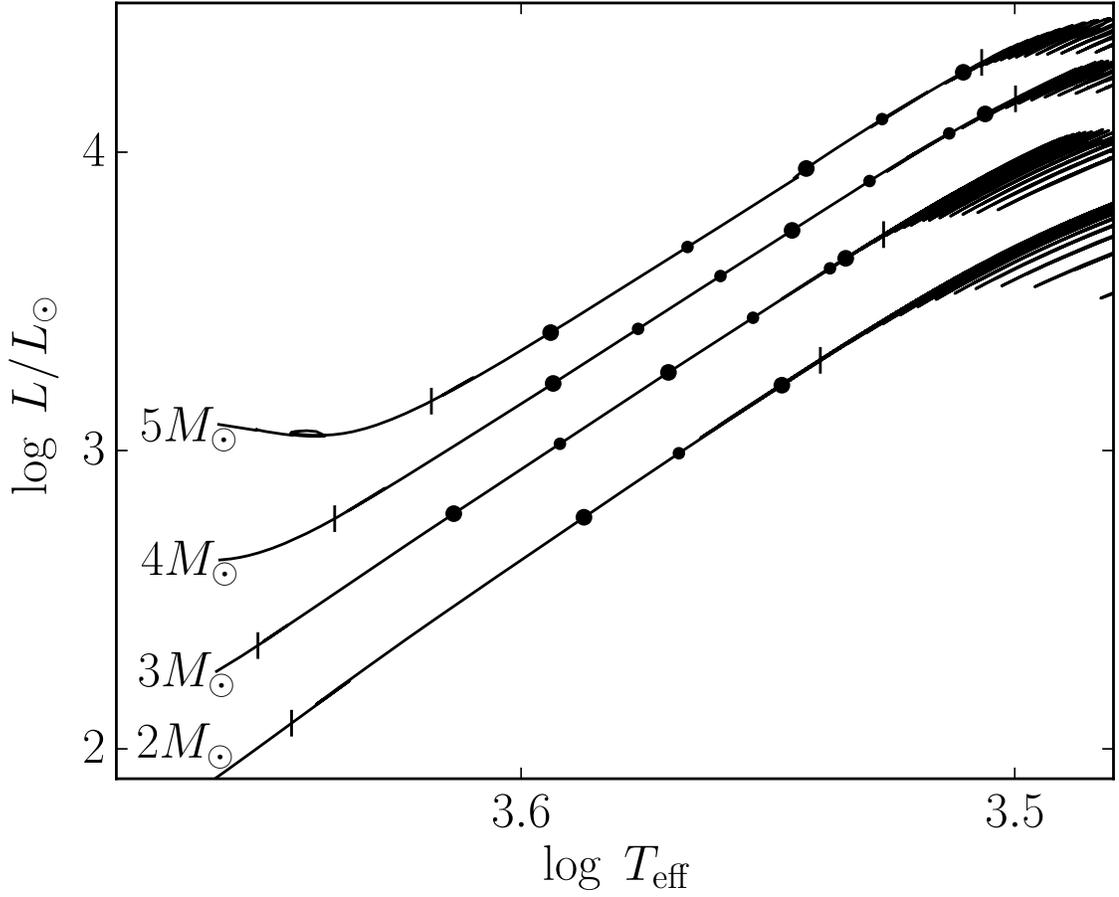}}
\caption{Evolutionary tracks of eAGB stars with initial masses
         $2M_\odot\le\mzams\le 5M_\odot$.
         Vertical ticks on each track indicate the end of core helium burning and
         the first thermal flash of the helium shell source.
         Filled circles show the hydrodynamic models.
         Larger circles indicate the models with parameters given in
         Table~\ref{table1}.}
\label{fig1}
\end{figure}

In this study we consider the eAGB stage as the time interval between the end
of helium core burning and the first thermal pulse of the helium shell source.
Evolutionary tracks and location of hydrodynamic models of eAGB stars
on the H--R diagram are shown in Fig.~\ref{fig1}.
Large filled circles indicate the models with parameters listed in
Table~\ref{table1}, where $M$, $L$ and $R$ are the mass,
the surface luminosity and the surface radius of the evolutionary
model, respectively.
The mean radius of the photosphere $\langle\rph\rangle$ and
the mean effective temperature $\langle\teff\rangle$
are determined for the limit cycle pulsations of the hydrodynamic model.
The pulsation period $\Pi$ and the pulsation constant $Q$ are expressed
in days.
The amplitude of the surface radial displacement in units of the initial
surface radius $\Delta R/R$, the amplitude of the surface velocity $\Delta U$
and the amplitude of bolometric light variations $\Delta M_\mathrm{bol}$
are given in last three columns.

\begin{figure}
\centerline{\includegraphics[width=16cm]{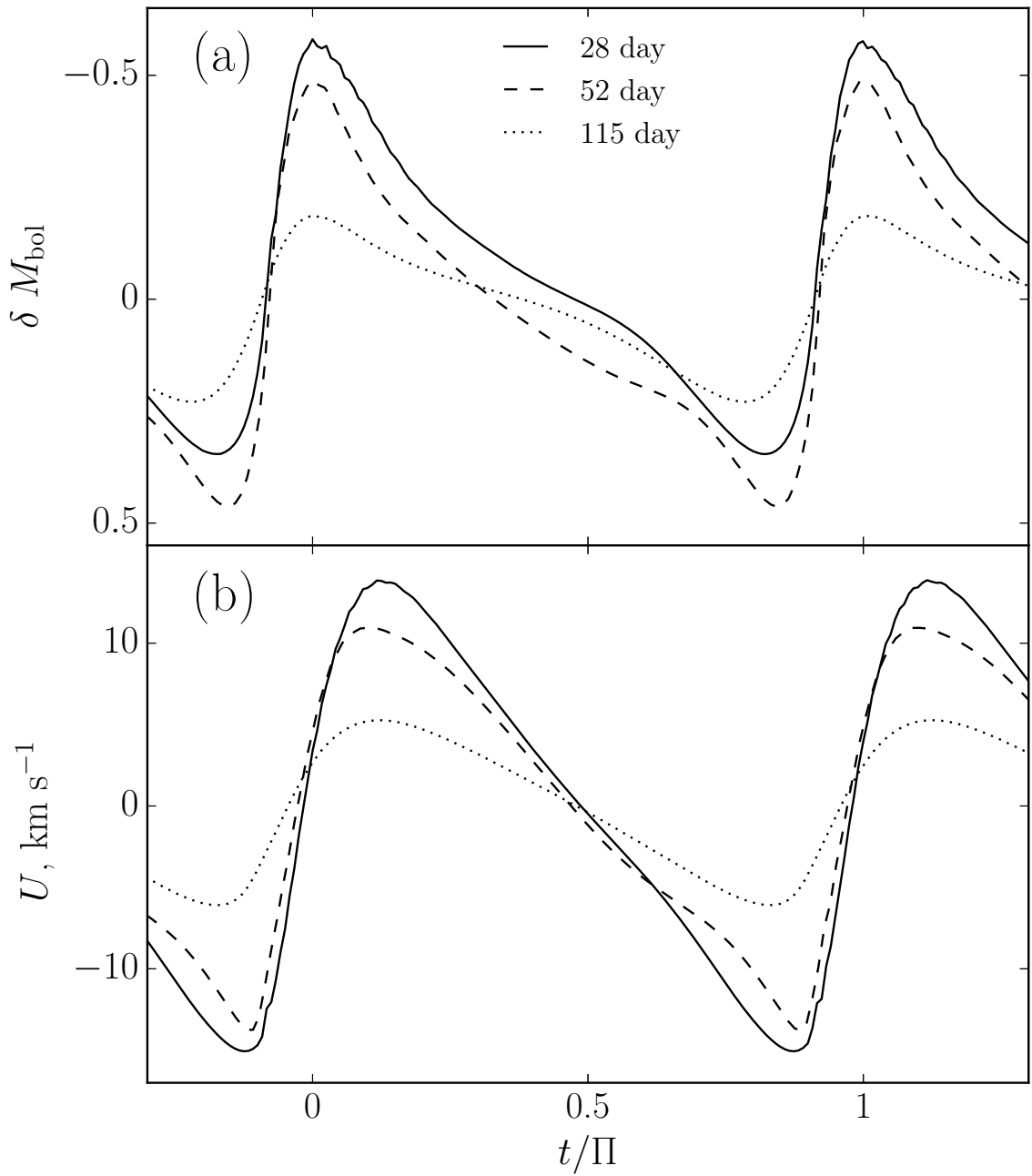}}
\caption{Temporal dependences of the bolometric light $\delta M_\mathrm{bol}$ (a) and
         the gas flow velocity at the outer boundary $U$ (b) for hydrodynamic models
         of eAGB stars with initial mass $\mzams=3M_\odot$ and pulsation periods
         $\Pi=28$, 52 and 115 day.}
\label{fig2}
\end{figure}

\begin{figure}
\centerline{\includegraphics[width=16cm]{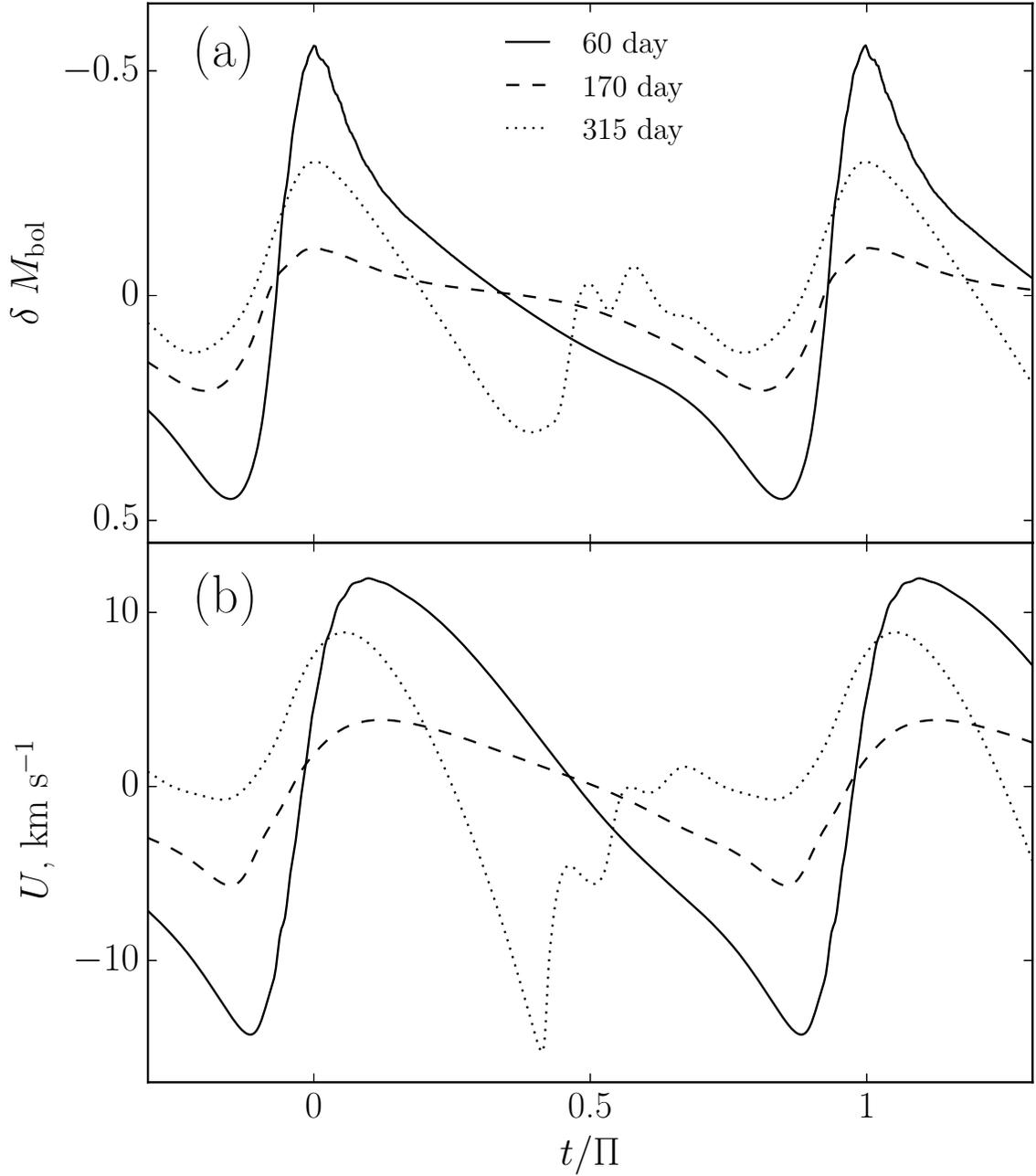}}
\caption{Same as Fig.~\ref{fig2} but for hydrodynamic models
         of eAGB stars with initial mass $\mzams=4M_\odot$ and pulsation periods
         $\Pi=60$, 170 and 315 day.}
\label{fig3}
\end{figure}

All eAGB red giants were found to be the fundamental mode pulsators with periods
from 15 to 400 day.
Figs.~\ref{fig2} and \ref{fig3} show the bolometric light and the surface velocity
curves of eAGB models of the evolutionary sequences with initial masses $\mzams=3M_\odot$
and $4M_\odot$, respectively.
The plots exhibit interesting features.
First, the amplitude of radial oscillations reduces as the star ascends the AGB.
This is due to the mass growth of the outer convection zone.
The pulsation amplitude reaches the minimum before the first thermal pulse.
Second, models with initial masses $\mzams\ge 4M_\odot$ and pulsation
periods $\Pi > 300$~d show the presence of the hump on the bolometric light
and the surface velocity curves.

Evolution of the eAGB red giant is accompanied by increase of the pulsation
period and the rate of period change but even in most luminous red giants
the rate of period change is insignificant:
$\dot\Pi/\Pi < 10^{-5}~\text{yr}^{-1}$.

\subsection{TP--AGB stars}

Dependence of pulsation properties of TP--AGB stars on the stellar mass $M$,
the surface luminosity $L$ and chemical composition is quite complicated.
In particular, hydrodynamic models of the evolutionary sequence $\mzams=3M_\odot$
exhibit fundamental mode pulsations near the minimum of the surface luminosity
and first overtone pulsations at the maximum luminosity (Fadeyev, 2016).
Hydrodynamic computations done in the present study show that fundamental mode
and first overtone pulsations appear in stars with initial mass $\mzams\le 3M_\odot$,
whereas more massive red giants pulsate only in the fundamental mode.

\begin{figure}
\centerline{\includegraphics[width=16cm]{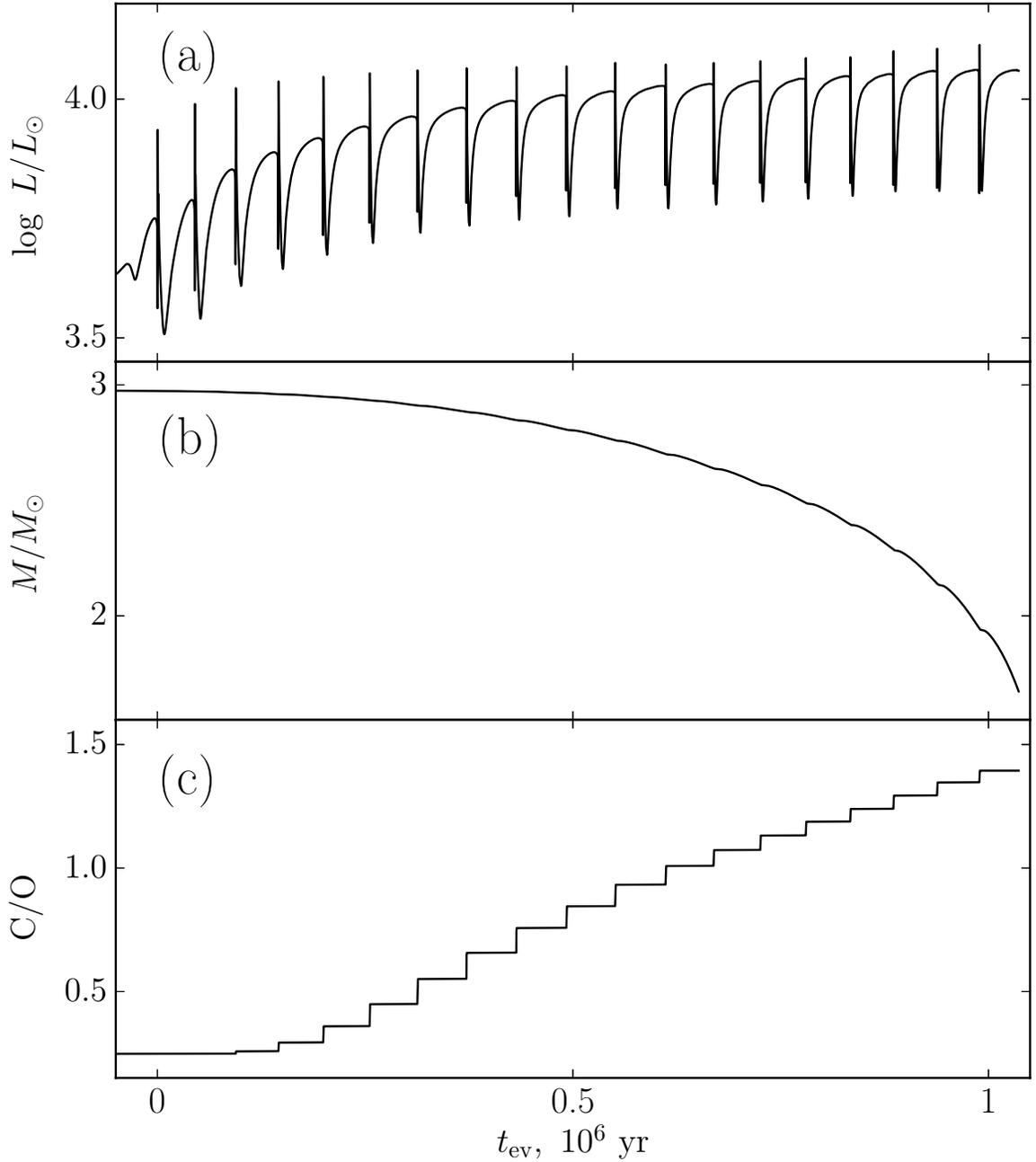}}
\caption{Temporal dependences of the surface luminosity $L$ (a),
         the stellar mass $M$ (b)
         and the carbon to oxygen mass fraction ratio C/O (c)
         of the evolutionary sequence $\mzams=3M_\odot$ within the evolutionary
         time interval including computed hydrodynamic models.}
\label{fig4}
\end{figure}

\textbf{Red giants with initial mass $\mathbf{\mzams=2M_\odot}$ and
$\mathbf{3M_\odot}$}.
Fig.~\ref{fig4} shows the plots of temporal dependences of the surface
luminosity $L$, the stellar  mass $M$ and the surface carbon to oxygen mass
fraction ratio C/O for stars of the evolutionary sequence $\mzams=3M_\odot$.
The time interval comprises 19 thermal pulses of the helium shell source.
For the sake of convenience hereinafter the evolutionary time $\tev$ is set
to zero at the first thermal flash.
The evolutionary sequence shown in Fig.~\ref{fig4} involves red giants of
spectral types M ($\mathrm{C}/\mathrm{O}<1$) and spectral types C
($\mathrm{C}/\mathrm{O}>1$).
Enrichment of outer stellar layers by products of nucleosynthesis is slower
in stars with lower initial mass.
The models of evolutionary sequence $\mzams=2M_\odot$ were computed for
the first 16 thermal flashes and the carbon to oxygen mass fraction was
$\mathrm{C}/\mathrm{O}\le 0.52$

\begin{figure}
\centerline{\includegraphics[width=16cm]{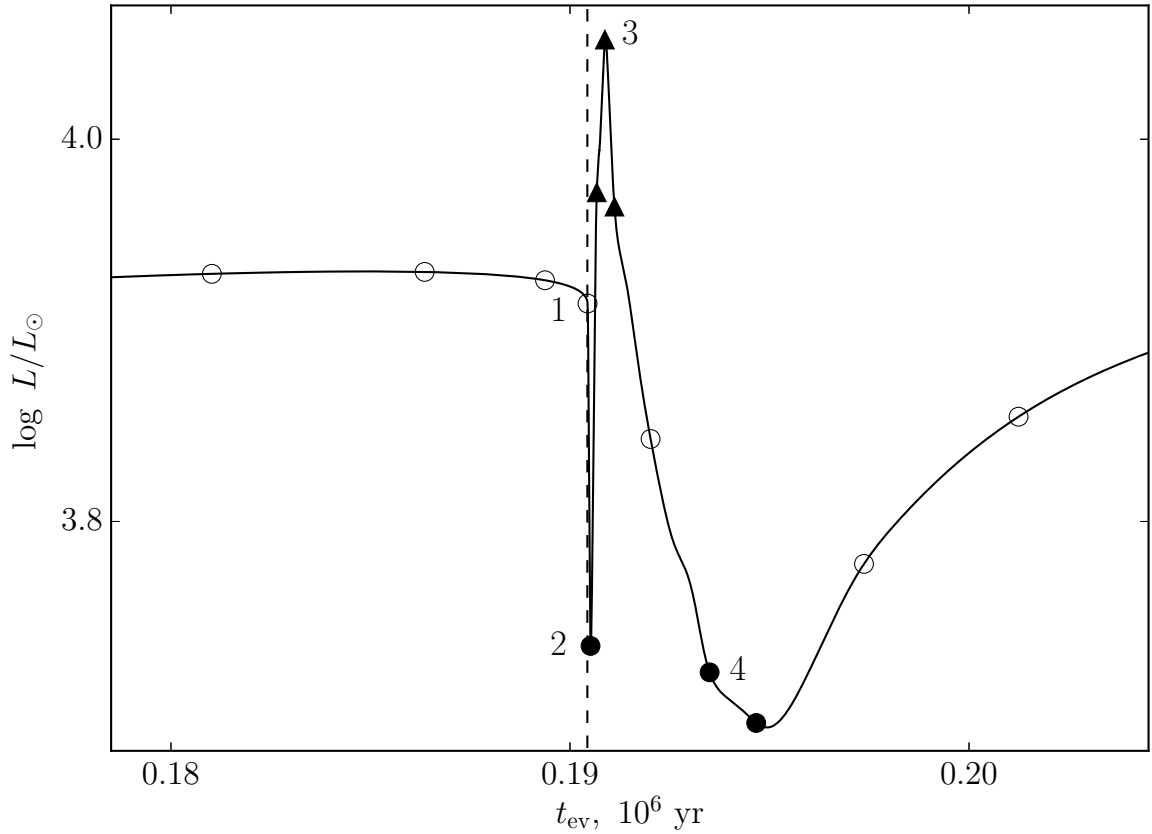}}
\caption{The surface luminosity (solid line) of stars of the evolutionary
         sequence $\mzams=3M_\odot$ during the fifth thermal pulse.
         The luminosity maximum of the helium shell source is marked by
         the vertical dashed line.
         Filled circles and filled triangles show hydrodynamic models of stars
         pulsating in the fundamental mode and the first overtone, respectively.
         Open circles show the pulsationally stable models with decaying
         oscillations.}
\label{fig5}
\end{figure}

In the present study we paid special attention to rapid temporal changes of stellar
luminosity and radius that occur after the luminosity maximum of the helium shell
source and are responsible for secular period changes observed in Myra--type variables.
Fig.~\ref{fig5} shows the surface luminosity of the red giant with initial mass $\mzams=3M_\odot$
as a function of evolutionary time during the fifth thermal pulse.
Filled circles and filled triangles indicate the mean luminosity of hydrodynamic models
with pulsations in the fundamental mode and in the first overtone.
Hydrodynamic models with decaying oscillations are marked by open circles.

The plot in Fig.~\ref{fig5} shows that during the thermal flash the stellar
oscillations change from the fundamental mode to the first overtone and
vice versa.
Moreover, the regions of pulsational instability in the fundamental mode and
in the first overtone are separated by the region of decaying oscillations.
Therefore, one may expect that together with secular period change some Mira--type
variables may exhibit secular change of the pulsation amplitude.
Indeed, simultaneous long--term changes of the period and the light amplitude
are observed in R Cen (Hawkins et al., 2001) and T UMi (Szatm\'ary et al., 2003;
Uttenthaler et al., 2011).
Unfortunately, in the present study detailed analysis of this interesting
feature of Miras is impossible because of the steady--state inner boundary
conditions implying the constant radius and the constant luminosity at
the bottom of the pulsating envelope.

\begin{figure}
\centerline{\includegraphics[width=16cm]{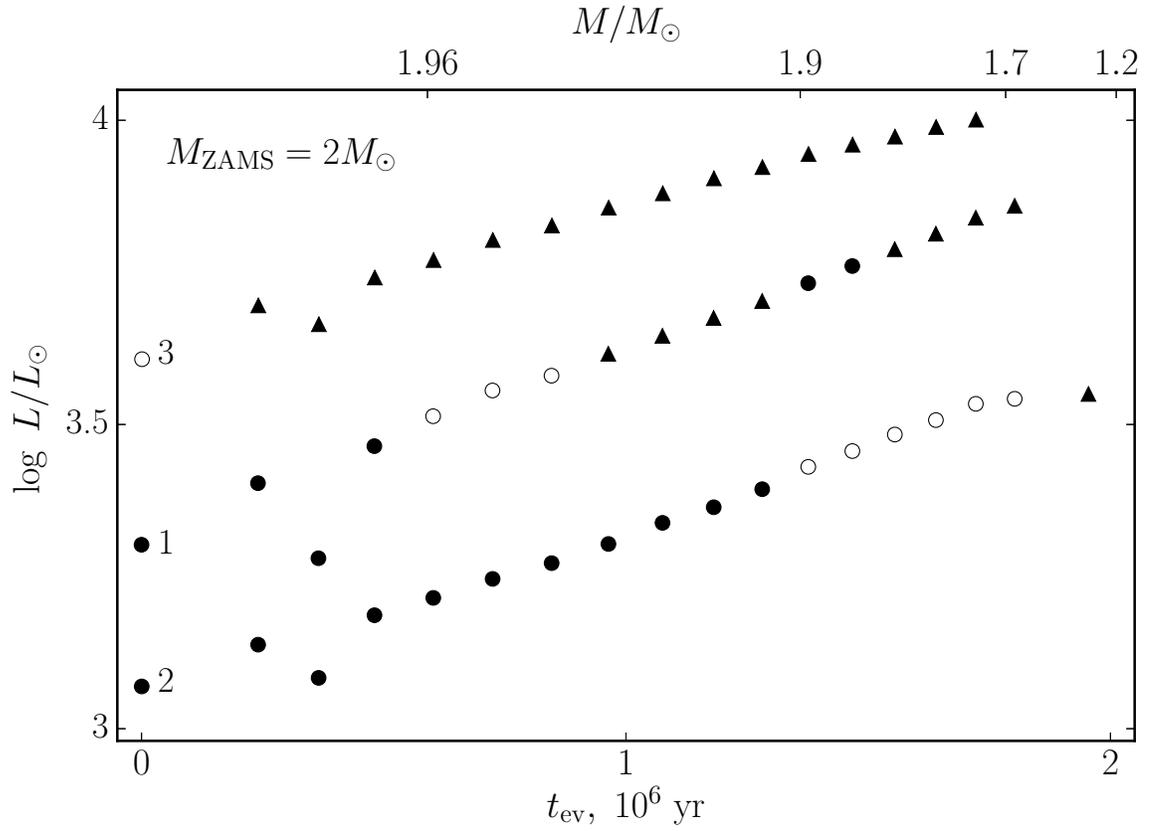}}
\caption{The luminosity of hydrodynamic models in points 1, 2 and 3 (see Fig.~\ref{fig5})
         as a function of the evolutionary time $\tev$ for the evolutionary
         sequence $\mzams=2M_\odot$.
         Designations of pulsation modes are same as in Fig.~\ref{fig5}.}
\label{fig6}
\end{figure}

\begin{figure}
\centerline{\includegraphics[width=16cm]{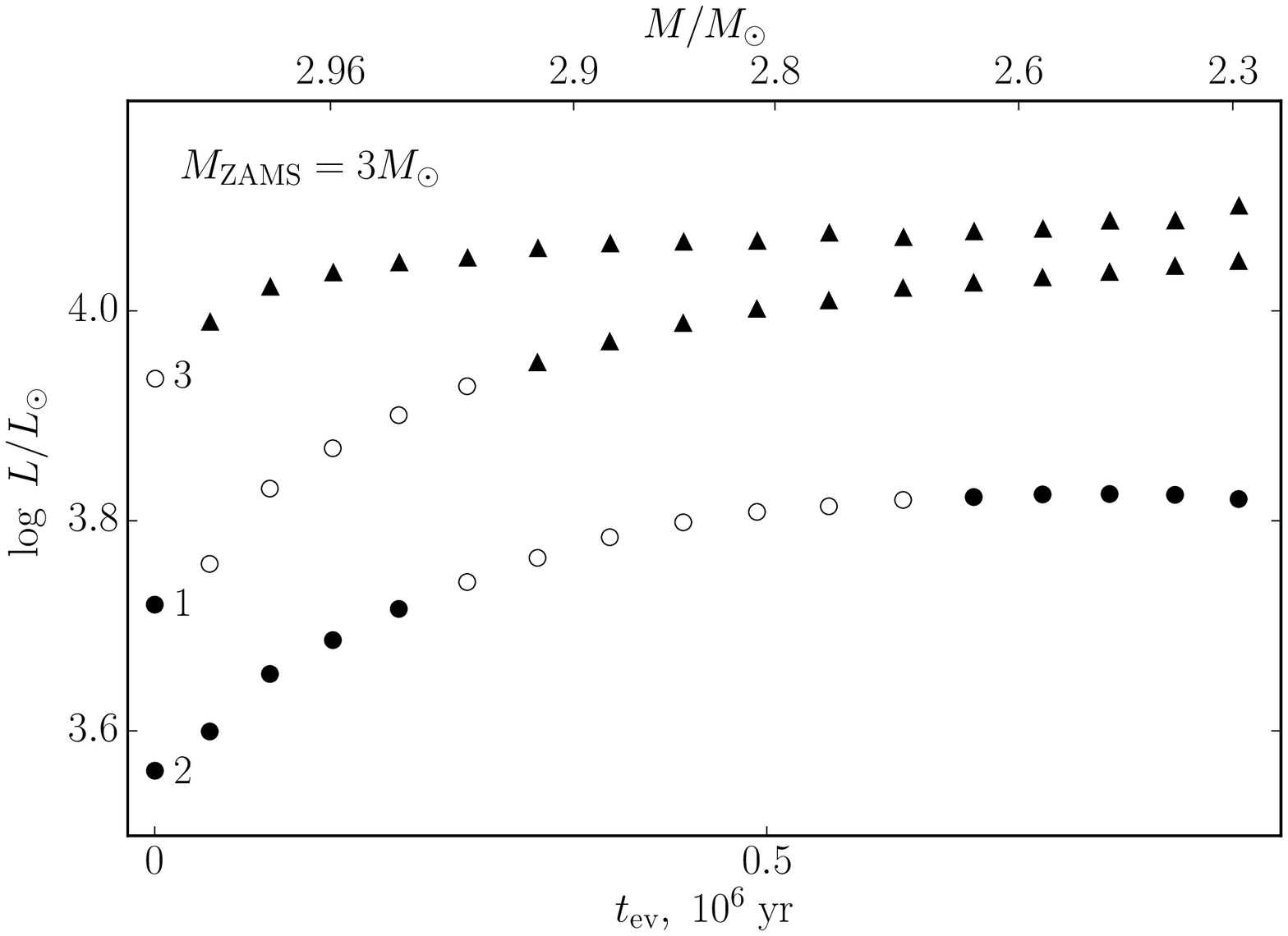}}
\caption{Same as Fig.~\ref{fig6} but for hydrodynamic models of the evolutionary sequence
         $\mzams=3M_\odot$.}
\label{fig7}
\end{figure}

As can be seen in Fig.~\ref{fig5}, the order of the oscillation mode depends on the
surface luminosity of the star.
However evolution of the red giant on the TP--AGB stage is accompanied by change of the
mass, luminosity and the chemical composition that are responsible for instability of
pulsation modes.
Changes in pulsation mode are shown in Figs.~\ref{fig6} and \ref{fig7} for hydrodynamic
models of evolutionary sequences $\mzams=2M_\odot$ and $\mzams=3M_\odot$.
Each thermal pulse is represented by three points with modes designatef as in
Fig.~\ref{fig5}.
The first point corresponds to the maximum of the luminosity of the helium shell source,
the second and the third points represent the minimum and the maximum surface
luminosity (see Fig.~\ref{fig5}).
Dependence of the pulsation mode on evolutionary changes of the stellar structure is clearly
seen in Fig.~\ref{fig7} due to transition from pulsational stability at the minimum surface
luminosity (the thermal pulses $6\le\itp\le 12$) to fundamental model pulsations for
$\itp > 12$.

\begin{figure}
\centerline{\includegraphics[width=16cm]{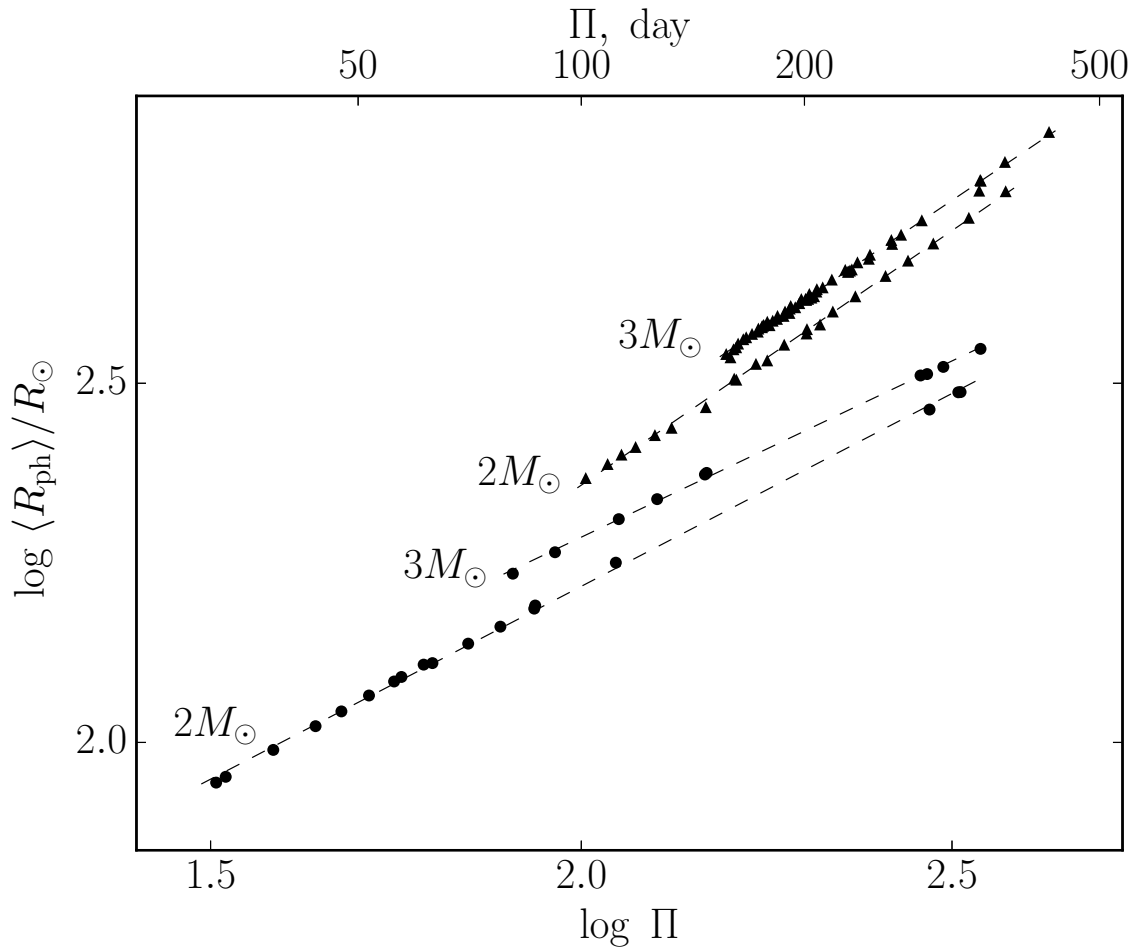}}
\caption{Period--radius dependences for the fundamental mode (filled circles)
         and the first overtone (filled triangles) pulsators of evolutionary sequences
         $\mzams=2M_\odot$ and $\mzams=3M_\odot$.
         Linear fits (\ref{pr2}) and (\ref{pr3}) are shown in dashed lines.}
\label{fig8}
\end{figure}

Typical values of the pulsation period of the fundamental mode and the first overtone
are illustrated by period--radius dependences in Fig.~\ref{fig8}.
Approximate linear fits of these dependences are given by
\begin{equation}
\label{pr2}
 \log \langle\rph\rangle = \left\{
\begin{array}{ll}
 1.161   +   0.527 \log\Pi  &  (\text{f--mode}) ,
   \\
 0.949   +   0.705 \log\Pi  &  (\text{1--st overtone})
\end{array}
\right.
\end{equation}
for $\mzams=2M_\odot$ and
\begin{equation}
\label{pr3}
 \log \langle\rph\rangle/R_\odot = \left\{
\begin{array}{ll}
 1.308 + 0.489 \log\Pi  &  (\text{f--mode}) ,
   \\
 1.015 + 0.696 \log\Pi  &  (\text{1--st overtone})
\end{array}
\right.
\end{equation}
for $\mzams=3M_\odot$.
The pulsation period $\Pi$ is expressed in days.

\textbf{Red giants with initial mass $\mathbf{\mzams=4M_\odot}$ and
$\mathbf{5M_\odot}$}.
Hydrodynamic models of TP--AGB stars with initial mass $\mzams=4M_\odot$ and  $5M_\odot$
were computed for evolutionary time intervals comprising first six thermal flashes.
All the models show pulsational instability in the fundamental mode with periods from
60 to $10^3$ day.
Approximate linear fits of the period--radius relation are written as
\begin{equation}
\label{pr45}
 \log \langle\rph\rangle/R_\odot = \left\{
\begin{array}{ll}
 1.342 + 0.496 \log\Pi,  &  (\mzams=4M_\odot) ,
   \\
 1.359 + 0.505 \log\Pi,  &  (\mzams=5M_\odot) .
\end{array}
\right.
\end{equation}

\section{The rate of period change}

In comparison of theoretical models with Mira--type variables of greatest
interest are the pulsation period $\Pi$ and the rate of period change $\dot\Pi$.
In the present study we approximately evaluated $\dot\Pi$
for three evolutionary time intervals with most rapid change of the
surface luminosity of the star.
Boundaries of the intervals are marked in Fig.~\ref{fig5} by numbers from 1
to 4.
The period change rate $\dot\Pi$ was evaluated from the difference of period
values at the interval boundaries.
In the case of decaying oscillations or pulsations in the different mode
on the edge of time interval we computed additional
hydrodynamic models and estimated the period change rate within the
narrower time interval.

\begin{figure}
\centerline{\includegraphics[width=16cm]{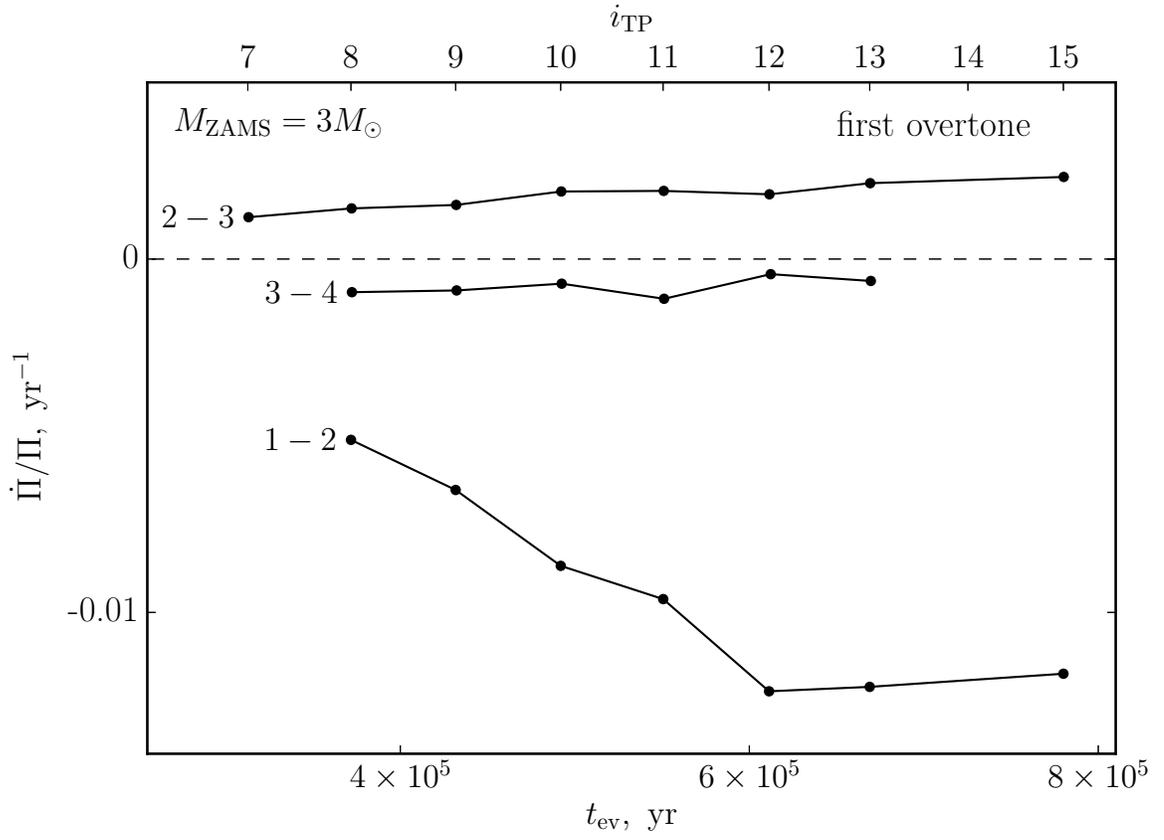}}
\caption{Mean values of the period change rate as a function of the
         evolutionary time $\tev$ for hydrodynamic models of the
         evolutionary sequence $\mzams=3M_\odot$.}
\label{fig9}
\end{figure}

\begin{figure}
\centerline{\includegraphics[width=16cm]{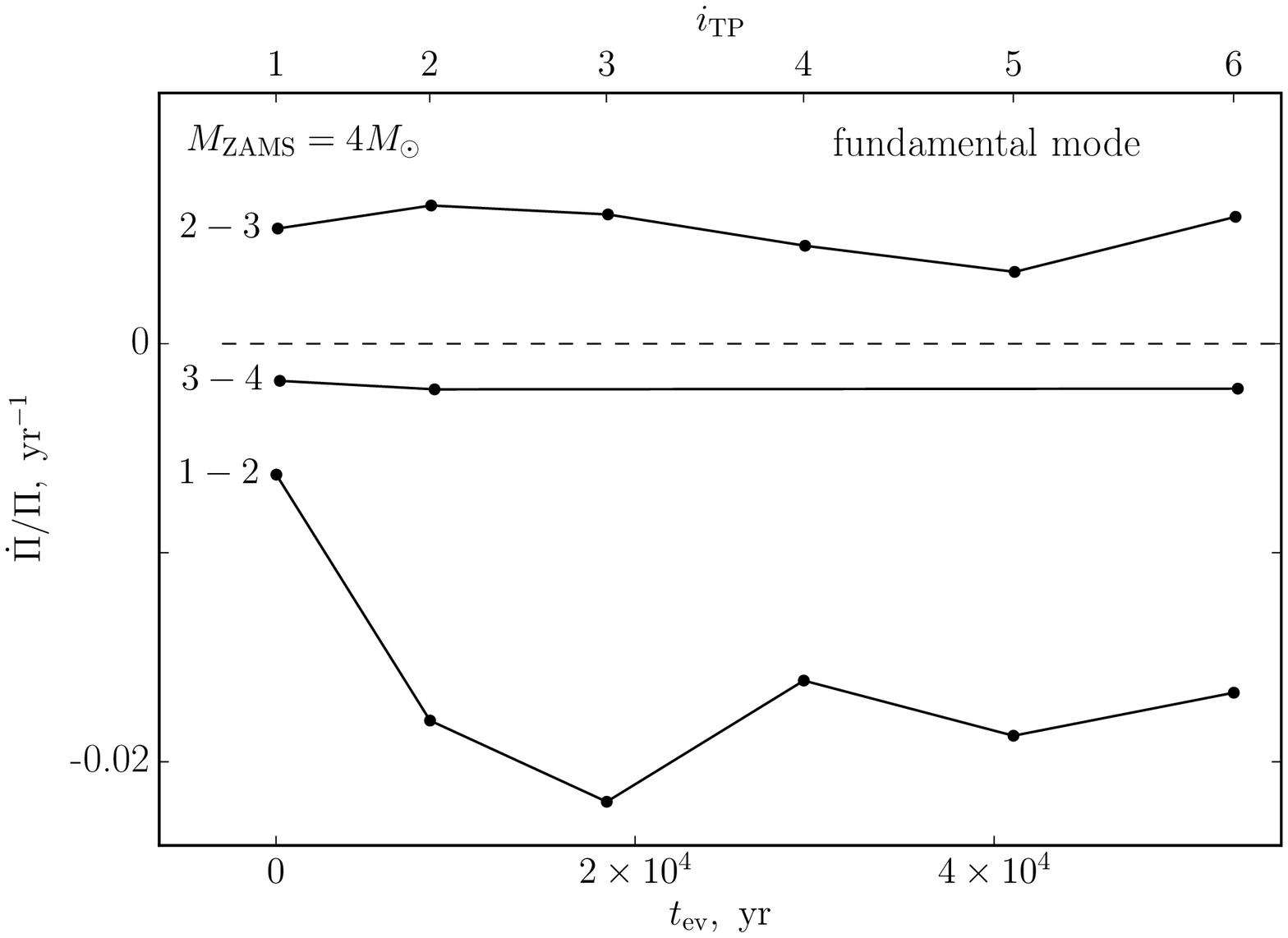}}
\caption{Same as Fig.~\ref{fig9} but for hydrodynamic models of
         the evolutionary sequence $\mzams=4M_\odot$.}
\label{fig10}
\end{figure}

Figs.~\ref{fig9} and \ref{fig10} show the approximate estimates of the period
change rate evaluated as
\begin{equation}
 \frac{\dot\Pi}{\Pi} = \frac{2}{\Delta\tev} \frac{\Pi_b - \Pi_a}{\Pi_a + \Pi_b} ,
\end{equation}
where $\Delta\tev = t_{\mathrm{ev},b} - t_{\mathrm{ev},a}$ -- is the time
interval and $\Pi_a$ and $\Pi_b$  are the pulsation periods at the edges of
the interval.
It should be noted that plots in Fig.~\ref{fig9} correspond to stellar
pulsations in the first overtone of the models $\mzams=3M_\odot$ whereas
plots in Fig.~\ref{fig10} represent the models $\mzams=4M_\odot$ pulsating
in the fundamental mode.

Mean values of $\dot\Pi/\Pi$ for four evolutionary sequences are given in
the last three columns of Table~\ref{table2}.
One should note that the mean value of the period change rate for the
evolutionary sequence $\mzams=2M_\odot$ might be underestimated due to
frequent mode switches (see Fig.~\ref{fig6}) as well as because of the
insufficient number of hydrodynamic models pulsating in the fundamental mode.
For models of the evolutionary sequence $\mzams=3M_\odot$ the mean values of
$\dot\Pi/\Pi$ were evaluated for thermal pulses $\itp>11$ when the period
change rate is roughly time--independent.
For evolutionary sequences $\mzams=4M_\odot$ and $\mzams=5M_\odot$
mean values of the period change rate were evaluated for $\itp>3$.

\section{conclusions}

This study was aimed to investigate the conditions corresponding to pulsations
of AGB stars in different radial modes.
We have shown that eAGB stars represent a homogeneous group of fundamental mode
pulsators with periods from 10 to 400 day, whereas TP--AGB stars show different
pulsation properties depending on the initial stellar mass.
In particular, TP--AGB stars with initial masses $4M_\odot\le\mzams\le 5M_\odot$
continue to pulsate in the fundamental mode with periods as long as $10^3$~day.
Less massive red giants ($2M_\odot\le\mzams\le3M_\odot$) evolving through
the TP--AGB phase show more intricate pulsation properties.
In particular, depending on the mass and the surface luminosity they pulsate
either in the fundamental mode or in the first overtone.
The upper limit of the first overtone period is
$\approx 300$ day for $\mzams=2M_\odot$ and $\approx 430$ day
for $\mzams=3M_\odot$.
Moreover, evolutionary changes of the stellar structure may lead to temporary
decay of radial oscillations.
Mira--type variables R Cen and T UMi seem to belong to such objects.

This study was aimed also to compare theoretical estimates of period change
rates in AGB stars with observations.
Pulsation periods of most investigated Miras T~UMi and LX~Cyg change with rates
$\dot\Pi/\Pi\approx -10^{-2}~\text{yr}^{-1}$ and
$\dot\Pi/\Pi\approx 7\times 10^{-3}~\text{yr}^{-1}$, respectively
(Templeton et al., 2005).
As we can see from Table~\ref{table2}, the theoretical estimates are in good
agreement with observations.

Of great importance is to evaluate the mass of TP--AGB stars using the stellar
evolution and stellar pulsation computations.
In particular, such an estimate can be obtained for the Mira T~UMi of spectral
type M5.5e (Keenan, 1966).
Before the early 1970s its pulsation period was $\Pi\approx 315$ day and showed
only small irregular changes.
However in 1976--1979 the period of T~UMi commenced abrupt decreasing with rate
evaluated from $-7.3\times 10^{-3}~\mathrm{yr}^{-1}$ to
$-1.2\times 10^{-2}~\mathrm{yr}^{-1}$
(J. G\'al, Szatm\'ary, 1995; Mattei, Foster, 1995; Szatm\'ary et al., 2003).
Therefore, the mass of T~Umi can be estimated from the evolutionary sequence of
hydrodynamic models provided that the surface luminosity and the surface radius
begin to decrease at the period 315 day.
First, the evolutionary sequences $\mzams\ge 4M_\odot$ should be excluded since
thermal pulses in these stars occur at pulsation periods $\Pi>500$ day.
In models of the evolutionary sequence $\mzams=3M_\odot$ the luminosity begins
to decrease at the period $\Pi\approx 300$ day during the 15--th thermal pulse
when the stellar mass is $M\approx 2.5M_\odot$.
However this model sequence should be excluded since the high carbon to oxygen
mass fraction ratio ($\mathrm{C}/\mathrm{O}=1.13$) corresponds to the spectral
type C.
The most appropriate estimate of the stellar mass can be obtained from the
evolutionary sequence $\mzams=2M_\odot$ for models of the 13--th thermal flash.
The surface luminosity begins to decrease at the fundamental mode period
$\Pi=325$ day and the carbon to oxygen mass fraction ratio is
$\mathrm{C}/\mathrm{O}=0.26$.
The mass and the luminosity of T~UMi are $M=1.9M_\odot$ and
$L=5.8\times 10^3L_\odot$, respectively.
It should be noted that these estimates are approximate and we give them
to illustrate the possibility to determine the fundamental parameters of
TP--AGB stars using the stellar evolution and stellar pulsation computations.
For more exact estimates of the mass and the surface luminosity of T~UMi we
have to use more detailed grids of stellar evolution and stellar pulsation
models.

\newpage
\section*{references}

\begin{enumerate}

\item T. Bloecker, Astron. Astrophys. \textbf{297}, 727 (1995).

\item E. B\"ohm--Vitense, Zeitschrift f\"ur Astrophys. \textbf{46}, 108 (1958).

\item M. Busso, R. Gallino, and G.J. Wasserburg, Ann. Rev. Astron. Astrophys. \textbf{37}, 239 (1999).

\item A.N. Cox and D.A. Ostlie, Astrophys. Space Sci. \textbf{210}, 311 (1993).

\item A.S. Eddington and S. Plakidis, MNRAS \textbf{90}, 65 (1929).

\item Yu.A. Fadeyev, MNRAS \textbf{449}, 1011 (2015).

\item Yu.A. Fadeyev, Pis'ma Astron. Zh. \textbf{42}, 731 (2016)
      [Astron. Lett. \textbf{42}, 665 (2016)].

\item J. G\'al, K. Szatm\'ary, Astron. Astrophys. \textbf{297}, 461 (1995).

\item C.A. Haniff, M. Scholz, P.G. Tuthill, MNRAS \textbf{276}, 640 (1995).

\item G. Hawkins, J.A. Mattei, G. Foster, Publ. Astron. Soc. Pacific \textbf{113}, 501 (2001).

\item F. Herwig, Astron. Astrophys. \text{360}, 952 (2000).

\item K.-H. Hofmann, M. Scholz and P.R. Wood, Astron. Astrophys. \textbf{339}, 846 (1998).

\item Y. Ita, T. Tanab\'e, N. Matsunaga, Y. Nakajima, C. Nagashima, T. Nagayama, D. Kato, M. Kurita, et al.,
      MNRAS \textbf{347}, 720 (2004).

\item D. Kamath, P.R. Wood, I. Soszy{\'n}ski,  and T. Lebzelter, MNRAS \textbf{408}, 522 (2010).

\item D.A. Keeley, Astrophys. J. \textbf{161}, 657 (1970).

\item P.C. Keenan, Astropys. J. Suppl. Ser. \textbf{13}, 333 (1966).

\item R. Kuhfu\ss, Astron. Astrophys. \textbf{160}, 116 (1986).

\item N. Langer,  M.F. El Eid and K.J. Fricke, Astron. Astrophys. \textbf{145}, 179 (1985).

\item D. Lorenz, T. Lebzelter, W. Nowotny, J. Telting,
                    F. Kerschbaum, H. Olofsson and H.E. Schwarz, Astron. Astrophys. \textbf{532}, A78 (2011).

\item J.A. Mattei, G. Foster, JAAVSO, \textbf{23}, 106 (1995).

\item P.W. Merrill, Astrophys. J. \text{116}, 21 (1952).

\item E.A. Olivier and P.R. Wood, MNRAS \textbf{362}, 1396 (2005).

\item B. Paxton, L. Bildsten, A. Dotter, F. Herwig, P. Lesaffre and F. Timmes,
                     Astropys. J. Suppl. Ser. \textbf{192}, 3 (2011).

\item B. Paxton, M. Cantiello,  P. Arras, L. Bildsten,
                     E.F. Brown, A. Dotter, C. Mankovich, M.H. Montgomery, et al.,
                     Astropys. J. Suppl. Ser. \textbf{208}, 4 (2013).

\item B. Paxton, P. Marchant, J. Schwab, E.B. Bauer,
                     L. Bildsten, M. Cantiello, L. Dessart, R. Farmer, et al.,
                     Astropys. J. Suppl. Ser. \textbf{220}, 15 (2015).

\item G. Perrin, S.T. Ridgway, B. Mennesson, W.D. Cotton, 
      J. Woillez, T. Verhoelst, P. Schuller, V. Coud{\'e} du Foresto, et al., Astron. Astrophys. \textbf{426}, 279 (2004).

\item J.R. Percy and T. Colivas, Publ. Astron. Soc. Pacific \textbf{111}, 94 (1999).

\item M. Perl and Y. Tuchman, Astrophys. J. \textbf{360}, 554 (1990).

\item D. Reimers, \textit{Problems in stellar atmospheres and envelopes}
      (Ed. B. Baschek, W.H. Kegel, G. Traving, New York: Springer-Verlag, 1975), p. 229.

\item R.L. Smith, W.K. Rose, Astrophys. J. \textbf{176}, 395 (1972).

\item K. Szatm\'ary, L.L. Kiss, and Zs. Bebesi, Astron. Astrophys. \textbf{398}, 277 (2003).

\item M.R. Templeton, J.A. Mattei, and L.A. Willson, Astron. J. \textbf{130}, 776 (2005).

\item Y. Tuchman, N. Sack, Z. Barkat, Astrophys. J. \textbf{219}, 183 (1978).

\item Y. Tuchman, N. Sack, Z. Barkat, Astrophys. J. \textbf{234}, 217 (1979).

\item P.G. Tuthill, C.A. Haniff, J.E. Baldwin, and M.W. Feast, MNRAS \textbf{266}, 745 (1994).

\item S. Uttenthaler, K. van Stiphout, K. Voet, et al., Astron. Astrophys. \textbf{531}, A88 (2011).

\item J. Weiner, Astrophys. J. \textbf{611}, L37 (2004).

\item P.R. Wood, Astrophys. J. \textbf{190}, 609 (1974).

\item P.R. Wood and D.M. Zarro, Astrophys. J. \textbf{247}, 247 (1981).

\item P.R. Wood and K.M. Sebo, MNRAS \textbf{282}, 958 (1996).

\item P.R. Wood, C. Alcock, R.A. Allsman, D. Alves, T.S. Axelrod, A.C. Becker, D.P. Bennett, K.H. Cook, et al.,
      \textit{Asymptotic Giant Branch Stars, IAU Symp. 191}
      (Ed. T. Le Bertre, A. L\`ebre, C. Waelkens, The Astronomical Society of the Pacific., 1999), p. 151.

\item P.R. Wood, E.A. Olivier and S.D. Kawaler, Astrophys. J. \textbf{604}, 800 (2004). 

\item A. Ya'ari, Y. Tuchman, Astrophys. J. \textbf{514}, L35 (1999).

\end{enumerate}

\newpage
\begin{table}
\caption{Hydrodynamic models of eAGB stars}
\label{table1}
\vskip 5pt
\begin{center}
 \begin{tabular}{ccrrrrrrr}
  \hline
  $\mzams/M_\odot$ & $M/M_\odot$ & $L/L_\odot$ & $R/R_\odot$ & $\Pi$, day & $Q$, day & $\Delta R/R$ & $\Delta U$, km/s & $\Delta M_\mathrm{bol}$ \\
  \hline                                                             
  2 & 1.965 &   598 &  54.6 &   15.5 & 0.0524 & 0.14 & 26 & 0.84 \\  
    & 1.965 &  1656 & 109.2 &   46.8 & 0.0572 & 0.17 & 21 & 0.85 \\  
  3 & 2.977 &   614 &  49.0 &   11.2 & 0.0543 & 0.10 & 24 & 0.56 \\  
    & 2.977 &  1827 & 103.2 &   34.8 & 0.0561 & 0.17 & 26 & 0.94 \\  
    & 2.975 &  4408 & 189.2 &  110.7 & 0.0658 & 0.13 & 12 & 0.46 \\  
  4 & 3.963 &  1678 &  88.9 &   24.1 & 0.0546 & 0.17 & 36 & 1.11 \\  
    & 3.962 &  5467 & 200.4 &   95.1 & 0.0630 & 0.16 & 19 & 0.83 \\  
    & 3.908 & 13432 & 376.1 &  375.8 & 0.0840 & 0.46 & 25 & 0.71 \\  
  5 & 4.938 &  2484 & 107.8 &   29.1 & 0.0545 & 0.19 & 43 & 1.32 \\  
    & 4.933 &  8816 & 257.9 &  134.8 & 0.0670 & 0.13 &  9 & 0.24 \\  
    & 4.870 & 18522 & 432.8 &  399.2 & 0.0831 & 0.54 & 30 & 0.96 \\  
  \hline
 \end{tabular}
\end{center}
\end{table}

\vskip 40pt
\begin{table}
 \caption{Mean values of the period change rate in TP--AGB stars}
 \label{table2}
 \begin{center}
 \begin{tabular}{ccccc}
  \hline
  $\mzams/M_\odot$   & mode &\multicolumn{3}{c}{$\dot\Pi/\Pi,\ \text{yr}^{-1}$} \\
                    &      & $1-2$  & $2-3$  & $3-4$  \\
  \hline
  2 & 0 & -0.002 &       &        \\
  3 & 1 & -0.010 & 0.002 & -0.001 \\
  4 & 0 & -0.016 & 0.005 & -0.002 \\
  5 & 0 & -0.020 & 0.005 &        \\
  \hline
 \end{tabular}
\end{center}
\end{table}
\clearpage

\newpage
\clearpage

\newpage
\clearpage

\newpage
\clearpage

\newpage
\clearpage

\newpage
\clearpage

\newpage
\clearpage

\newpage
\clearpage

\newpage
\clearpage

\newpage
\clearpage

\newpage
\clearpage

\end{document}